# A lateral-type spin-photodiode based on Fe/x-AlO$_x$/p-InGaAs junctions with a refracting-facet side window

Ronel Christian Roca\*, Nozomi Nishizawa, Kazuhiro Nishibayashi, and Hiro Munekata\*

*Laboratory for Future Interdisciplinary Research of Science and Technology, Tokyo Institute of Technology, Yokohama 226-8503, Japan*

\*E-mail: roca.r.aa@m.titech.ac.jp

A lateral-type spin-photodiode having a refracting facet on a side edge of the device is proposed and demonstrated at room temperature. The light shed horizontally on the side of the device is refracted and introduced directly into a thin InGaAs active layer under the spin-detecting Fe contact in which spin-polarized carriers are generated and injected into the Fe contact through a crystalline AlO$_x$ tunnel barrier. Experiments have been carried out with a circular polarization spectrometry set up, through which helicity-dependent photocurrent component, $\Delta I$, is obtained with the conversion efficiency $F \approx 0.4$ %, where $F$ is the ratio between $\Delta I$ and total photocurrent $I_{\text{ph}}$. This value is the highest reported so far for pure lateral-type spin-photodiodes. It is discussed through analysis with a model consisting of drift-diffusion and quantum tunneling equations that a factor that limits the $F$ value is unoccupied spin-polarized density-of-states of Fe in energy region into which spin-polarized electrons in a semiconductor are injected.

# 1. Introduction

Spin-optoelectronics, or spin-photonics, is an emerging sub-field of spintronics that aims at adding spin-based functionality on conventional optoelectronic devices.[1,2] While most of the effort is focused on the development of circularly polarized light (CPL) sources such as spin light-emitting-diodes (spin LED)[3,4] and spin lasers,[5] the development of a CPL detector, the spin photodiode (spin-PD), is just as important. The spin-PD does not require external optical delay modulators and thus simple, and furthermore, it can convert helicity-based signals into electrical signals with wide bandwidth transmission, which is essential in spin-based optical communication.[6,7] Up to now, most studies on spin-PDs involves vertical-type devices.[6-12]

The vertical-type devices are advantageous for coupling with external optical components due to their large active area and compatibility with optical interconnects.[6] However, this configuration restricts the choice of magnetic metals to those that exhibit perpendicular magnetic anisotropy (PMA). Additionally, since light is incident onto the top surface of the spin-PD, unless a window is formed in the magnetic metal contact, magnetic circular dichroism (MCD) due to the magnetic contact is inherently present in their measurements,[12] and results in spurious detection of helicity of light. On the other hand, for a lateral-type device on which light is incident onto the side edge of the device, the PMA requirement is relaxed, and MCD contribution can be suppressed to a great extent.[13] Most importantly, the lateral-type devices are better suited for intra-chip device-to-device optical communications, as well as monolithic integrated circuits which contain multiple optoelectronic devices (emitters and detectors) in a single chip. So far, a few studies have been reported for lateral-type spin-PDs, incorporating oblique angle incidence.[13-15]

In the previous studies, the figure of merit $F = \Delta I / I_{ph}$ has been introduced,[8,14] in which $\Delta I = I_{ph}(\sigma^+) - I_{ph}(\sigma^-)$ and $I_{ph} = [I_{ph}(\sigma^+) - I_{ph}(\sigma^-)] / 2$. Here, $\sigma^+$ and $\sigma^-$ represent right and left CPL, respectively. In the present study, we call $F$ as the helicity conversion efficiency. Due to recent advances, $F$ as high as 5 % has been reported for vertical type devices at room temperature (RT),[8,9] whereas $F \approx 0.1$ % has been reported for a pure edge-illuminated lateral-type device at RT.[13] More recently, an improved value of $F > 1$ % has been reported for lateral-type devices with oblique angle illumination.[14,15] In particular, the experiment with the incident angle of 60°[14] has suggested that direct illumination on the cleaved edge may involve some effects that degrade the process of spin transport near the cleaved edge, such as surface recombination and trapping of photo-generated carriers/spins.

In this work, we propose a lateral-type spin-PD having a refracting facet side window.[16,17] Shown in Fig. 1 (a) is a schematic of the proposed spin-PD. The light illuminated directly on the side of the device is bent by the refracting facet and sent directly onto the active layer below a spin-detecting magnetic contact where spin-polarized carriers are generated and transported to the magnetic contact through a tunnel barrier. This configuration is expected to circumvent the problem associated with the cleaved edge. Moreover, due to the illumination geometry, it is expected that the contribution of MCD is much less than those of purely vertical and oblique angle geometries.[14,15] Besides, the refracting-facet structure is also expected to exhibit improved high speed response.[17] A Schottky-barrier-type junction with a p-type InGaAs layer is adopted on the basis of consideration that the transport length of photo-generated, spin-polarized electrons should be kept as short as possible; in other words, the layer that converts light helicity into electron spins should be placed as close as possible to the spin-detecting metal contact. In addition, a crystalline γ-like $AlO_x$ tunnel barrier is inserted in between the metal and the p-InGaAs layers, aiming at suppressing the annihilation of photo-generated electrons around the interface.[18] Namely, the tunnel barrier in the present device is expected to take roles of suppressing the non-radiative recombination as well as the interface chemical reaction between ferromagnet metal and semiconductor layers. Furthermore, the tunnel barrier may also improve spin detection efficiency by avoiding the conduction mismatch problem in semiconductor-based spintronic devices.[19-21]

Characterization of fabricated, refracted-facet spin-PDs has been carried out at RT with a circular polarization spectrometry set up, through which helicity-dependent photocurrent component with the experimental helicity conversion efficiency $F \approx 0.4\%$ is demonstrated without the application of an external magnetic field. This value is the highest reported so far for pure-lateral-type spin-PDs. A model calculation is also presented, and the mechanism that limits the $F$ value is discussed together with possible solutions.

## 2. Experiment

Crystalline oxide - semiconductor structures were grown by molecular beam epitaxy. A 280-nm-thick epitaxial p-GaAs:Be ($N_A \approx 5\times10^{17}$ cm$^{-3}$) buffer layer was first grown at the substrate temperature of $T_s = 580°C$ on a p-GaAs:Zn (001) substrate ($N_A \approx 10^{19}$ cm$^{-3}$). This was followed by the growth of a 400-nm or 40-nm thick p-In$_{0.05}$Ga$_{0.95}$As:Be ($N_A \approx 5\times10^{17}$ cm$^{-3}$) active layer at $T_s = 510°C$. The wafers were then cooled to $T_s = 80°C$ or lower at which a seed Al epitaxial layer was grown on top of the p-InGaAs:Be layer. The Al epilayer was oxidized in dry air atmosphere at room temperature to yield a 1-nm crystalline γ-like $AlO_x$

($\gamma$-AlO$_x$) layer. Details of the growth of $\gamma$-AlO$_x$ can be found elsewhere.[22]

The 480-μm-wide, magnetic stripe contacts consisting of Fe (100 nm)/Ti (10 nm)/Au (20 nm) layers were formed on $\gamma$-AlO$_x$/p-semiconductor wafers by standard vacuum deposition and photolithography techniques; Fe and Ti layers were deposited by e-beam evaporation, whereas a Au layer by resistive evaporation. The bottom contacts were formed by resistive evaporation of a 40-nm thick indium layer on the back side of the wafer.

The refracting facet was then fabricated by an anisotropic wet chemical etching using a H$_2$SO$_4$:H$_2$O$_2$:H$_2$O (1:8:1) solution. It is known that the etch rate is fastest for the [11-1] direction and slowest for the [111] direction.[23] Prior to the etching, magnetic contact stripes were completely covered by 520-μm-wide stripe photoresist. A more detailed description of the device fabrication can be found elsewhere.[24] The refracting-facet spin-PD thus prepared has a facet angle of $\theta_{facet} \approx 68°$ with respect to the wafer plane with ± 2° variation [Fig. 1 (b)]. The oblique angle of a light beam impinging on a p-InGaAs layer is 19.5°. A light beam of 100 μm width along the $z$ direction results in the 290-μm long, photo-excitation area along the $x$ direction [Fig. 1 (a)], which is narrower than the width of a magnetic contact. The estimated inclination of light intensity along the $x$ axis over the width of 290 μm is around 20 %.

Fabricated refracted facet wafers were annealed at 230 °C for 1 hour in the N$_2$ atmosphere, and were cleaved into individual spin-PD chips having approximately 1 mm square with a magnetic contact dimension of 480μm × 1 mm. The stripes' long axis was aligned along the [1-10] direction, as shown in Fig. 1 (b) and (c).

Shown in Fig. 2 (a) is the optical measurement setup. A light beam is shed on a refracting face widow along the GaAs [110] axis (the $x$ axis in the figure). A Tsunami Ti:Sapphire pulse laser, with pulse width of ≈ 150 fs and repetition rate ≈ 80 MHz, was used in order to vary the central excitation wavelength from 840 (1.48 eV) to 930 nm (1.33 eV). Figure 2 (b) depicts photographically the way how a spin-PD was mounted on a sample stage. A linearly polarized light beam from the laser was converted into a CPL beam by using a Glan-Thompson linear polarizer (LP) and a quarter-wave plate (QWP). Helicity switching of the beam between left ($\sigma^-$) and right ($\sigma^+$) CPL was carried out by manually rotating the QWP. The CPL beam of radium approximately 450 μm was shed on the refracting facet using a lens of focal length $f$ = 30 cm and $NA$ = 0.033. Measurements were carried out using the lock-in technique with a mechanical chopper operating at 400 Hz. The average light power impinged on the chip was adjusted to have a constant value of 3.6 mW (28 nJ/cm$^2$ per pulse) for all measurements. The corresponding peak photon flux per pulse is around $1.3 \times 10^{11}$

photons/cm$^2$ at the wavelength λ = 900 nm. Shown in Fig. 2 (c) is the dark and illuminated *I-V* curves of a fabricated spin-PD with a 400-nm-thick InGaAs active layer. The *I-V* curve shifts down when a light beam (λ = 900 nm, average power 3.6 mW) is shed on the refracting facet window, yielding a photocurrent of around 16 µA at 0 V (short circuit condition). The fill factor[25] of 0.31 is estimated, which suggests the presence of a finite amount of leak current in the spin-PD. For the CPL-specified photocurrent (CPL-ph) measurements, a 500-Ω load resistor is connected in series to the spin-PD. A load line based on the total resistance,[11] $R_{total} = R_{load} + R_{SPD}$ ($R_{SPD}$ ~ 500 Ω), is also depicted in Fig. 2 (c), through which a photocurrent of ≈ 8 µA with a photo-voltage of ≈ 10 mV is expected.

The CPL-ph measurements were carried out by switching the helicity of a light beam several times between σ$^-$ to σ$^+$ while keeping the remanent magnetization direction unchanged. The values of $I_{ph}(\sigma^+)$ and $I_{ph}(\sigma^-)$ were determined by time-averaging the value of measured CPL-ph for typically 50 s. Measurements were also carried out under the opposite remanent magnetization (*M*) direction. The shape anisotropy in the plane is small, as exemplified by the *M-H* curves taken across {the [110] axis, Fig. 2 (d)} and along {the [1-10] axis, Fig. 2 (e)} the long side of a stripe. Therefore, nearly full magnetization value is kept at the remanent *M* state after switching the original *M* by applying an opposite external field (400 Oe). The CPL-ph is measured in the form of lock-in output voltage across the resistor. See appendix section for the relation between pulse excitation and time-averaged photocurrent.

## 3. Results and discussion

### 3.1 CPL specified photocurrent

Shown in Fig. 3 (a) is the temporal profile of the measured photocurrent $I_{ph}$ for two different remanent magnetization states, in which the magnetization vector pointing towards the light source (+Rem) and the other with magnetization pointing reversely (−Rem) for a refracting-facet spin-PD with *d* = 400 nm. No bias voltage was applied on the tested spin-PD. For the profile measured with the +Rem state (blue profile), the $I_{ph}$ value increases (decreases) when the helicity of the incident laser beam is changed from σ$^+$ to σ$^-$ (σ$^-$ to σ$^+$) polarization. When the measurement is carried out with the −Rem state (red profile), the relative change upon switching the light helicity is reversed. These results indicate that the presence of helicity-dependent photocurrent in the proposed device configuration.

A slight drift, in the order of less than 1%, is observed in particular for the −Rem data. We infer that this comes from the slight but unavoidable mechanical drift of a mirror (not shown) used to steer the incident laser beam onto the device. We estimate, on the basis of the distance between the device and the mirror (85 cm), that the angular drift in the order of $10^{-6}$ deg. causes approximately 1% change in the measured photocurrent. We tried to eliminate this effect by designing the run sequence with an odd number of measurement windows (i.e. $\sigma^+ \to \sigma^- \to \sigma^+ \to \sigma^- \to \sigma^+$) such that the drift is averaged out. A similar run sequence, made for the same purpose, has been utilized in Ref. 1.

Shown in Fig. 3 (b) is the wavelength dependencies of photocurrent $I_{ph}$ and helicity conversion efficiency $F$. No external bias voltage is applied. The wavelength was varied from 840 to 930 nm while keeping the incidence power at 3.6 mW. It can be seen that the $I_{ph}$ and $F$ are both maximized at $\lambda$ = 900 nm ($hv$ = 1.38 eV), indicating that photogeneration of spin-polarized carriers occurs primarily in a p-InGaAs layer but not in a p-GaAs layer and a substrate. In detail, both $I_{ph}$ and $F$ decrease with increasing the wavelength, which is consistent with reduced absorbance toward the fundamental absorption edge of $In_{0.05}Ga_{0.95}As$ ($E_g$ = 1.35 eV, $\lambda$ = 920 nm). The $I_{ph}$ value significantly dropped at $\lambda$ = 930 nm. Both $I_{ph}$ and $F$ also decrease with decreasing the wavelength from $\lambda$ = 900 to 880 nm and shorter ($hv \geq$ 1.41 eV). In this wavelength region, light absorption starts taking place in the p-GaAs region, which reduces the number of photogenerated electrons in the p-InGaAs layer. The reduction rate of the $F$ value towards the shorter wavelengths is more severe compared to that of $I_{ph}$. This is because of the rather short spin diffusion length ($\lambda_{spin} \approx$ 1.3 µm) compared to the relatively long minority carrier diffusion length of 21µm in p-GaAs.[14] The $F$ value obtained at $\lambda$ = 900 nm is $F \approx$ 0.4%, which is around four times larger than that obtained from the lateral spin-PD without a refracting-facet window.[14]

The width of the depletion region increases when the reverse (negative) bias is applied on the spin-PD, under which photogenerated spin-polarized electrons are expected to be transported more efficiently towards the magnetic contact [inset of Fig. 4 (b)]. Shown in Fig. 4 (a) are the plots of $I_{ph}$ and $\Delta I$ as a function of the applied bias voltage. It is clearly seen that both $I_{ph}$ and $\Delta I$ increase with increasing the reverse (negative) bias voltage. When a positive bias is applied, both $I_{ph}$ and $\Delta I$ decrease, as expected. On the other hand, as shown in Fig. 4 (b), the $F$ value remains nearly constant with different bias voltages. We infer that the observed bias independence of $F$ is because the spin relaxation length is not affected by the voltage within the limit of the present work ($|V| <$ 1 V ).[26, 27]

Note that the InGaAs active layer has a lattice mismatch to GaAs of $\Delta a/a_{GaAs}$ = 0.38%, which gives rise to the critical layer thickness of $\approx$ 40 nm.[28] Namely, the tested spin-PD with an active layer of 400 nm is inferred to have threading dislocations throughout the InGaAs layer and misfit dislocations at the InGaAs/GaAs interface. We therefore tested anther refracting-facet spin-PD with InGaAs thickness $d$ = 40 nm. Results of CPL-ph experiments for the 40-nm spin-PD clearly exhibit the presence of $\Delta I$ signals with less noise, as shown in Fig. 5 (a). However, dramatic increase in $F$ value is not observed [Fig. 5 (b)]. This fact suggests that the $F$ values in the present devices are not primarily limited by misfit dislocations and associated crystalline defects.

### 3.2 Analysis based on spin-charge transport

A model for spin-charge transport consisting of drift-diffusion[29-31] and quantum tunneling[32, 33] transports is developed in order to seek ways to further improve the $F$ value. As shown in Fig. 6, a source light beam enters from the backside of an InGaAs layer, and reflected back at the metal-oxide-semiconductor interface. These two processes, depicted by the illumination with first ($\Phi_1$) and second ($\Phi_2$) beams, yield photo-generated electrons whose population is represented by the quasi Fermi level $E_F^*$ that has a downward gradient towards the edge of the depletion region ($z$ = 41 nm): $E_F^*$ is spin-split when a CPL beam is shed as exemplified in inset Fig. 6. A change in the helicity of $\Phi_2$ due to magnetic circular dichroism (MCD) of a Fe electrode is as small as $5.0 \times 10^{-3}$, and thus negligible.[14] Diffusion driven by the gradient of $E_F^*$ is a predominant transport process in the neutral region ($z \geq 41$ nm), whereas a drift process participates in the transport in the depletion region (0 < z < 41 nm) in which an electric field of $E_{dp}$ = $1.4 \times 10^5$ V/cm is present. In this region, the charge transport time, $t_{dp} \approx w/(\mu_e E_{dp})$, is reduced down to around $10^{-14}$ s which is much shorter than spin relaxation time. Furthermore, the charge/spin transport direction is parallel to the direction of $E_{dp}$. Because of these reasons, we assume no degradation in spin polarization during the transport across the depletion region. Finally, electrons/spins that reach at the γ-AlO$_x$/p-InGaAs interface are injected into a Fe electrode through an oxide tunnel barrier.

One-dimensional drift-diffusion equations shown by Eqs. (1) and (2) with $\Delta n = \Delta n^\uparrow + \Delta n^\downarrow$ and $\Delta s = \Delta n^\uparrow - \Delta n^\downarrow$ are utilized to simulate the transport in a semiconductor. We neglect the inclination of light intensity along the $x$ axis for simplicity.

$$\frac{\partial \Delta n}{\partial t} = D_e \frac{\partial^2}{\partial z^2} \Delta n + \mu_e E_{dp} \frac{\partial}{\partial z} \Delta n - \frac{\Delta n}{\tau_{rec}^*} + G, \qquad (1)$$

$$\frac{\partial \Delta s(P)}{\partial t} = D_e \frac{\partial^2}{\partial z^2}\Delta s(P) + \mu_e E_{dp}\frac{\partial}{\partial z}\Delta s(P) - \frac{\Delta s(P)}{\tau_{spin}^*} + G_{spin}(P). \quad (2)$$

Here, $D_e$ ($\approx 62$ cm$^2$/s) is the electron diffusion coefficient, $\mu_e$ the electron mobility [$\approx 2400$ cm$^2$/(V·s)],[34] $\tau_{rec}^* = \left(\frac{1}{\tau_{rec}} + \frac{1}{\tau_{nr}}\right)^{-1}$ the effective minority carrier recombination time incorporating the bulk minority carrier radiative recombination time $\tau_{rec}$ ($\approx 7.2 \times 10^{-8}$ s)[35,36] and the non-radiative recombination time $\tau_{nr}$, and $\tau_{spin}^* = \left(\frac{1}{\tau_{spin}} + \frac{1}{\tau_{rec}^*}\right)^{-1}$ with the bulk, spin relaxation time $\tau_{spin}$ ($\approx 2.3 \times 10^{-10}$ s).[36] $G$ is the time-averaged carrier generation rate, expressed by $G(z) = \alpha \Phi(z)/sin(\theta)$ with $\Phi(z) = \Phi_1 + \Phi_2$, $\theta \approx 20°$ the incident angle of the light with respect to the $x$-axis, whereas $\alpha \approx 10^4$ the representative absorption coefficient above the fundamental absorption edge. $G_{spin}$ is the spin generation rate that has the relation $G_{spin} = P \cdot (0.5 \cdot G)$ with $P = \pm 1$ for $\sigma^\pm$ on the basis of the optical selection rule.[37] Note that $\tau_{nr}$ is a sample-dependent unknown parameter in these equations. Setting a drift-diffusion photocurrent $J_{d-d}$ at $z = 0$ (the boundary condition), relation between $J_{d-d}$ and photogenerated electrons $\Delta n$ is calculated with various $\tau_{nr}$ by the conventional finite difference method (FDM).[38] The size of a finite segment is set $\Delta z = 1$ nm. The calculated $J_{d-d}$ is then compared with the experimental photocurrent to find the probable $\tau_{nr}$ value. Results of calculations with $\tau_{nr} = 1.7 \times 10^{-12}$ s, the likely value in our tested device, are exemplified in Supplemental Material. Energy position of the nominal quasi-Fermi level at $z = 0$ is shown in Fig. 6 by a dot placed on a vertical line of γ-AlO$_x$ / p-InGaAs interface, which indicates no significant charge/spin accumulation at the interface.

Quantum tunneling equation, as shown by Eq. (3), is utilized to simulate the transport across the oxide barrier. It is important that current conservation condition is imposed across the entire region of metal-oxide-semiconductor: namely, $J_{d-d} = J_{tunnel}$ at $z = 0$.

$$J_{tunnel} = A \int \{D_{sc} f_{sc}(E - E_F^*) D_m [1 - f_m(E - E_F)]$$
$$- D_m f_m(E - E_F) D_{sc}[1 - f_{sc}(E - E_F^*)]\} T(E) dE. \quad (3)$$

Here, $A$ is constant with the unit of cm$^{-4}$·C·eV·s$^{-1}$, $D$ the density-of-states (DOS), $f(E)$ the Fermi distribution function, $E_F^*$ the quasi-Fermi level for electrons in a semiconductor, and $T$ the tunneling probability; the subscripts $sc$ and $m$ represent the semiconductor (SC) and metal (M) sides of the junction, respectively. We estimate $T = 0.052 \pm 0.033$ on the basis of the WBK approximation[39,40] with a barrier height of $1.55 \pm 0.1$ eV [22] and barrier thickness of $1 \pm 0.2$ nm. Owing to a small number of time-averaged photo-generated carriers (see Supplemental Material), the range of integral in the Eq. (3) can be reduced down to the bottom of the conduction band $E_C$ using the effective DOS of the SC conduction band, $N_C$, and Boltzmann distribution: namely,

$$J_{tunnel} \approx AD_m(E_C)T(E_C) \int D_{sc} f_{sc}(E - E_F^*)\, dE$$
$$\approx AD_m(E_C)T(E_C)N_C \exp\left(\frac{E_C - E_F^*}{k_B T}\right) = AD_m(E_C)T(E_C)\Delta n. \quad (4)$$

Shown in Fig. 7 (a) are the plots of $J_{d-d}$ as a function of $\Delta n$ for three different $\tau_{rec}^*$ values, $7.2 \times 10^{-8}$, $1.0 \times 10^{-10}$, and $1.7 \times 10^{-12}$ s with $\tau_{nr}$ values of $\tau_{nr} \to \infty$, $1.0 \times 10^{-10}$, and $1.7 \times 10^{-12}$ s, respectively. A plot of $J_{tunnel}$ is also presented in the figure. Solutions of Eqs. (1) and (4) for different $\tau_{rec}^*$ values are obtained at the intersection of $J_{d-d}$ and $J_{tunnel}$ curves. Results of calculations are re-plotted in the form of the *J-T* relation in Fig 7 (b), and compared with the experimental $J_{exp} \sim 10$ mA/cm². Within the limit of $T = 0.052 \pm 0.033$, we are able to find $\tau_{rec}^* \approx \tau_{nr} = 1.7 \times 10^{-12}$ s for the 400-nm spin-PD. This value is close to that of the non-radiative recombination near the metal-semiconductor interface.[41,42]

Let us finally examine the spin transport in tunneling, which is expressed as:

$$J^{\uparrow\downarrow} \approx AD_m^{\uparrow\downarrow}T \int \{D_{sc} f_{sc}(E - E_F^{*\uparrow\downarrow})\, dE \approx AD_m^{\uparrow\downarrow}T\Delta n^{\uparrow\downarrow}. \quad (5)$$

The total current is expressed as:

$$J = J^{\uparrow} + J^{\downarrow} = \left[AT\frac{(D_m^{\uparrow} + D_m^{\downarrow})}{2}\Delta n\right] + \left[AT\frac{(D_m^{\uparrow} - D_m^{\downarrow})}{2}\Delta s\right] = J_0 + J_\sigma. \quad (6)$$

The first term $J_0$ indicates purely charge current and takes the form similar to that of Eq. (4) with $D_m = \frac{(D_m^{\uparrow} + D_m^{\downarrow})}{2}$, whereas the second term $J_\sigma$ represents the helicity-dependent component of the photocurrent. From Eq. (6), it is straight forward that the helicity-dependent photocurrent can be expressed as:

$$\Delta J = J(\sigma^+) - J(\sigma^-) = [J_0 + J_\sigma(\sigma^+)] - [J_0 + J_\sigma(\sigma^-)]$$
$$= AT(D_m^{\uparrow} - D_m^{\downarrow})|\Delta s(z = 0)|. \quad (7)$$

We now recognize that the difference in unoccupied DOS between the spin-up and -down in Fe, $\Delta D = D_m^{\uparrow} - D_m^{\downarrow}$, is an important quantity that determines the efficiency of spin-PDs. In the calculation, DOS for Fe in the energy range equivalent to the conduction band edge of a InGaAs is assumed to be around $1.4 \times 10^{23}$ cm⁻³ eV⁻¹,[43] and the coefficient *A* around $1.4 \times 10^{-30}$ cm⁻⁴·C·eV·s⁻¹.

Shown in Fig. 7 (c) is calculated helicity dependent photocurrent $\Delta J$ as a function of $\Delta D$ for $\tau_{rec}^* \approx 1.7 \times 10^{-12}$ s. Referring the experimentally measured $\Delta J_{exp} \approx 0.04$ mA/cm$^2$, we are able to extract the $\Delta D$ value of around $1.2 \times 10^{21}$ cm$^{-3}$ eV$^{-1}$ with $|\Delta s(z=0)| = 4.8 \times 10^8$ cm$^{-3}$. The $\Delta D$ value thus obtained amounts to 0.85 % out of the total DOS of Fe, which is quite small in view of a ferromagnet. We point out, however, that spin cross-over in the density-of-states may occur in the energy region in which spin polarized electrons are injected: it is around 1 eV above the Fermi level.[43-46] This argument is not yet conclusive, since we ignore other experimental factors that may give rise to reduction of $\Delta J$; namely, degradation of spin polarization due to poor magnetic quality of Fe near the interface[47] and/or presence of a spin-independent, leak current. Assuming the ideal fill factor of 1 and Fe DOS spin polarization of $\Delta D/D_m \approx 0.4$, $F > 10\%$ is expected.

Our analysis for spin-PD utilizing minority carrier injection suggests that one of the most direct ways to improve $\Delta J$ is to increase $\Delta D$ by using a ferromagnet whose empty DOS have half-metallic character at the energy range sufficiently higher than the Fermi level. For p-GaAs based spin-PD, Co$_2$FeMnSi quaternary alloy would be one of the candidates, since this material may have relatively high spin polarization (P > 0.8) at the energy range that is 1 eV higher than the Fermi level.[48] Another possible scenario is to suppress a process of non-radiative recombination near the metal-oxide-semiconductor junction. For this approach, improvement in the crystalline quality of an ultrathin γ-AlO$_x$ tunnel barrier should be pursued. This approach will increase both $\Delta J$ and $J_{ph}$, but the ratio $F$ would not be improved significantly. It is also interesting to look into a tunnel barrier that exhibits spin-filter effect (e.g. MgO). In this scenario, tunneling probability is not equal ($T^\uparrow \neq T^\downarrow$) between two different spins, and thus an enhanced $\Delta J$ value is expected.

## 4. Conclusions

We have proposed a lateral-type spin-photodiode incorporating a refracting facet window on the side wall of the diodes, and have presented results of experiments at room temperature together with model calculations. Experimental results show helicity-dependent photocurrent component with helicity conversion efficiency $F \approx 0.4\%$, which is the highest reported so far for the pure, lateral-type spin-photodiodes. Through model calculations, small spin polarized DOS of Fe is suggested as one of the possible origins for the relatively small $F$ value. Possible directions for future studies have also been suggested.

## Acknowledgments

The authors would like to acknowledge support in part by the Advanced Photon Science Alliance from the Ministry of Education, Culture, Sports, Science and Technology (MEXT) and the Matching Planner Program from Japan Science and Technology Agency (JY290153). R.C.R. acknowledges the scholarship from MEXT. N.N. acknowledges the support by the Japan Society for Promotion of Science (JSPS) Grant-in-Aid (KAKENHI) No. JP17K14104 and a research granted from the Murata Science Foundation.

# Appendix

Schematically shown in Fig. A.1 are the expected temporal profiles of incident pulse photon flux $\Phi(t)$, decay function of photogenerated electrons $D(t)$, and resulting photocurrent $J(t)$. The light pulse arrives at time $t = t_0$, whereas $J(t)$ starts increasing as the pulse arrives, reaches its maximum at $t = t_0$, and then decays. The decay process is described by the decay function $D(t) = \exp(-t/\tau_{rec}{}^*)$ for $t \geq 0$, in which $\tau_{rec}{}^*$ is the lifetime of electrons. The relationship between $J(t)$ and $\Phi(t)$ can be represented mathematically using the convolution operation[49] and is described by

$$J(t) = eC \cdot (\Phi * D) = eC \cdot \int_0^\infty \Phi(t_1) \cdot D(t - t_1) \, dt_1. \tag{A.1}$$

Here, $e$ is the magnitude of the electron charge, $C$ is a constant in units of cm$^2$ and $t_1$ is a dummy variable for integration (note that the functions $J$, $\Phi$, and $D$ are only defined for $t \geq 0$). The value of $C$ can be determined experimentally via time-resolved photocurrent measurements but this is beyond the scope of the present work. It is shown in the discussion section that $\tau_{rec}{}^*$ is in the order of $\approx 10^{-12}$ s, whereas the time interval between pulses $T_{pp}$ is about 12.5 ns ($\approx 10^{-8}$ s) and the pulse width is $\tau_p \approx 150$ fs ($\approx 10^{-13}$ s). Since $T_{pp} \gg \tau_{eff}$, we can treat each pulse as isolated.

In experiment, the measurable output is the time average of $J(t)$, namely $\langle J \rangle$:

$$\langle J \rangle = \frac{1}{T_{pp}} \cdot \int_0^{T_{pp}} J(t) \, dt, \tag{A.2}$$

where we take the average during one period $T_{pp}$. Substituting Eq. (A.1) into (A.2) yields:

$$\langle J \rangle = \frac{1}{T_{pp}} \cdot \int_0^{T_{pp}} eC \cdot (\Phi * D) \, dt \approx \frac{eC}{T_{pp}} \cdot \int_0^\infty (\Phi * D)(t) \, dt. \tag{A.3}$$

Here, the assumption that the pulses are isolated allows us to change the limits of the integration to infinity. Note that although $J(t)$ is time-varying, $\langle J \rangle$ is time-independent (quasi-steady-state) and can directly be used to estimate the steady-state solution of Eq. (1). Furthermore, the form of Eq. (A.3) allows us to utilize the integration property of convolutions, which yields:

$$\langle J \rangle \approx \frac{eC}{T_{pp}} \cdot \int_0^\infty (\Phi * D) \, dt = \frac{eC}{T_{pp}} \cdot \left[ \int_0^\infty \Phi(t) \, dt \right] \cdot \left[ \int_0^\infty D(t) \, dt \right]$$

$$\approx e \left[ C \cdot \int_0^\infty D(t) \, dt \right] \cdot \left[ \frac{1}{T_{pp}} \cdot \int_0^{T_{pp}} \Phi(t) \, dt \right] = e\eta \cdot \langle \Phi \rangle. \tag{A.4}$$

Here, we again utilize the assumption of isolated pulses to change the limits of integration and gather all the terms in the first square bracket into a constant $\eta$, which is just effective quantum efficiency in the present work, whereas the second bracket is simply the time average of the photon flux, which is equal to the time-average value used in the experiment.

The value of quantum efficiency $\eta$ is directly affected by the effective lifetime $\tau_{rec}^*$. From the experimental data, we estimate $\eta$ to be around $\approx 3\%$. In other words, we are able to directly correlate the time-average values of $J(t)$ with $\Phi(t)$ even for the case of pulse excitation, using the steady-state solution of Eq. (1). Note that a similar analysis can be done for the solution of Eq. (2).

**Figure Captions**

**Figure 1** (a) Schematic cross section of a refracting-facet spin-PD. Arrows in a Fe layer represent direction of remanent magnetization which is parallel to the *x* axis. A light beam is shed horizontally from the left side on the refracting facet. (b) Side view of the cleaved edge of a fabricated refracting-facet spin-PD observed by scanning electron microscopy (SEM). The facet angle $\theta_{facet}$ is approximately 68° with respect to the *x* axis that is parallel to the GaAs [110] axis. Facet height and etch depth are approximately 70 and 112 μm, respectively. Red arrows represent beam directions in a spin PD. (c) Bird's-eye-view of the same spin-PD observed by SEM. Blue and red arrows represent the GaAs [1-10] and [110] axes, respectively.

**Figure 2** (a) Schematic illustration of circular polarization (CP) spectrometry setup. CP is generated by passing a light beam from a Ti:Sapphire laser through a linear polarizer (LP) and a quarter-wave plate (QWP). The CP laser beam is focused onto the sample with a spot size of radius ≈ 450 μm using a lens with the focal length *f* = 30 cm and NA = 0.033. (b) A picture of a tested spin-PD that is fixed on a copper sample holder by firmly pressing it with a Cu metal finger. (c) *I-V* curves of a tested spin-PD having a 400-nm thick InGaAs layer in the dark (blue) and under the illumination (red) with a light beam of the wavelength λ = 900 nm. Straight line represents a load resistance line (green). *M-H* hysteresis curves obtained from a tested spin-PD with (d) magnetic fields applied along (d) the GaAs [110] axis (the *x*-axis) and (e) the GaAs [1-10] axis (the *y* axis). Magnetic characteristics are nearly same for both curves.

**Figure 3** (a) Temporal profiles of CPL-specified photocurrent measured with a light beam of wavelength λ = 900 nm for a spin PD comprising a 400-nm thick InGaAs layer. + Rem and − Rem indicate magnetization vector point towards + and −*x* axis, respectively. Data obtained with + / − Rem state are separated vertically for graphic clarity. Dashed lines (black) are drawn for eye guides. (b) Photocurrent $I_{ph}$ and helicity conversion efficiency *F* as a function of wavelength of an impinged light beam. Vertical dashed lines (black) crossing λ = 870 and 920 nm denote, respectively, the band gap energy of GaAs and InGaAs.

**Figure 4** (a) Plots of measured photocurrent $I_{ph}$ (blue) and helicity-dependent photocurrent $\Delta I$ (red) as a function of applied bias voltage. (b) A plot of measured $F$ values as a function of applied voltage. Inset of (b) shows band edge profiles with application of reverse bias.

**Figure 5** (a) Temporal profiles of CPL-specified photocurrent for spin-PD incorporating 40-nm thick InGaAs layer qmeasured either with + or − Rem state. The wavelength of a CPL beam is λ = 900 nm. Data are separated vertically for graphic clarity. Horizontal dashed lines (black) are drawn for eye guides. (b) A plot of measured $F$ values as a function of applied voltage for a 40-nm spin-PD.

**Figure 6** Schematic band edge profiles in spin-PD. The labels C.B. and V.B. stand for conduction and valence band edge, respectively. Thickness of light absorbing p-InGaAs layer ($N_A = 5 \times 10^{17}$ cm$^{-3}$) is set at 400 nm. Diffusion potential across the Fe/AlO$_x$/p-InGaAs junction is around 0.6 V which is distributed between the AlO$_x$ tunnel barrier ($V_B \approx 0.02$ V) and the Schottky depletion layer ($V_{Sch} \approx 0.58$ V). The width of the depletion region is $w \approx 41$ nm with the Schottky barrier height of $\varphi_{Sch} \approx 0.58$ eV. Tunnel barrier height is $\varphi_B \approx 1.55$ eV from the conduction band edge. $E_F$ and $E_F^*$ represent the Fermi level in the dark and quasi-Fermi level under the illumination with light, respectively. Two half-parabolas at the most left side of figure represent schematically spin-polarized DOS of Fe. A dark arrow represents the first light beam $\Phi_1$ entering an InGaAs layer, whereas a light arrow does the second beam $\Phi_2$ that is reflected back at the γ-AlO$_x$ / p-InGaAs interface. Graded intensities of $\Phi_1$ and $\Phi_2$ are schematically shown by broken lines. Inset shows a split of the Fermi level, $E_F^{*\uparrow}$ and $E_F^{*\downarrow}$, when CPL is shed on spin-PD. Illumination with intensity of 3.6 mW results in $\Delta E_F^* = E_F^{*\uparrow} - E_F^{*\downarrow} = 11$ meV at the p-InGaAs/p-GaAs interface.[29] A dot on the vertical line of γ-AlO$_x$ / p-InGaAs interface indicates the position of nominal $E_F^*$ representing the number of photogenerated electrons at $z = 0$.

**Figure 7** (a) Calculated drift-diffusion current $J_{d-d}$ and tunneling current $J_{tunnel}$ as a function of photo-generated electron concentration $\Delta n$ at $z = 0$. For $J_{d-d}$, three different effective recombination lifetimes, $\tau_{rec}^* \approx 7.2 \times 10^{-8}$ s (blue diamonds), $1.0 \times 10^{-10}$ s (green triangles), and $1.7 \times 10^{-12}$ s (red squares) are assumed. Self-consistent solutions are given by the intersections of $J_{d-d}$ and $J_{tunnel}$ curves. (b) Calculated photocurrent $J$ as a function of the tunneling rate $T$ with four different $\tau_{rec}^*$ values; $7.2 \times 10^{-8}$ s (blue diamonds), $1.0 \times 10^{-10}$ s (red squares), $1.7 \times 10^{-12}$ s (green triangles) and $1.0 \times 10^{-13}$ s (purple crosses). Horizontal dashed line expresses the experimental photocurrent value of $J \approx 10$ mA/cm$^2$. A hatched region in orange represents tunneling probability $T = 0.052 \pm 0.033$. (c) Calculated helicity-dependent photocurrent $\Delta J$ as a function of the difference in DOS $\Delta D$ between spin-up and -down bands. Horizontal dashed line indicates the experimental helicity-dependent photocurrent value $\Delta J_{exp} \approx 0.04$ mA/cm$^2$, whereas vertical dashed line indicates extracted $\Delta D$ value of around $1.2 \times 10^{21}$ cm$^{-3}$ eV$^{-1}$.

**Figure A.1** Schematic illustrations of temporal profiles of pulsed photon flux $\Phi(t)$, the decay function $D(t)$, and resulting photocurrent $J(t)$. The form for $J(t)$ represents the convolution of $\Phi(t)$ and $D(t)$. The light pulse arrives at time $t = t_0$.

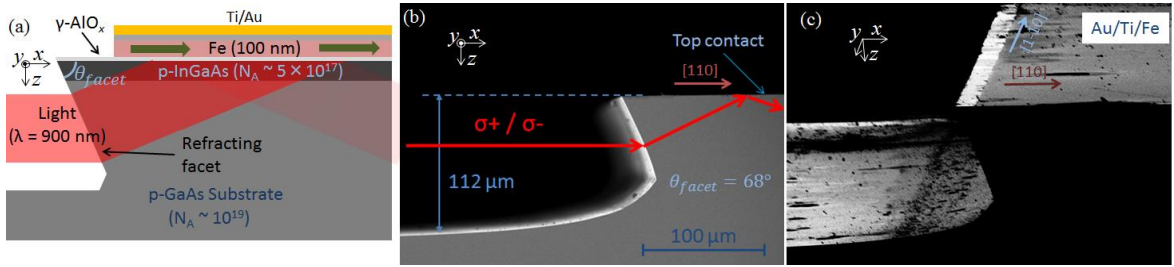

Fig. 1. (Color online)
20

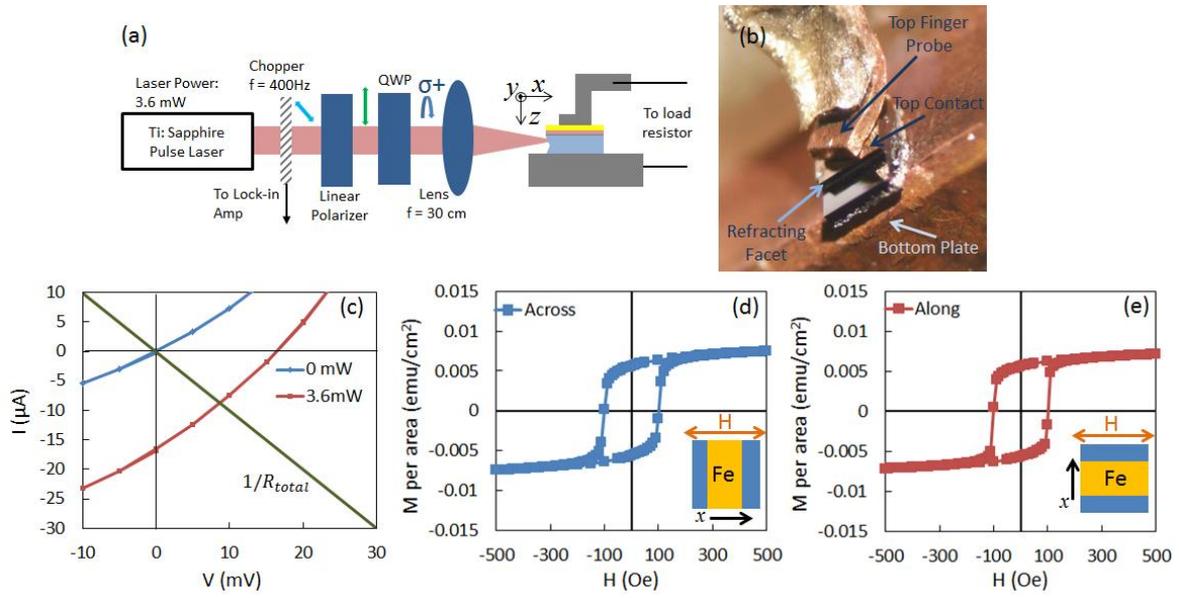

Fig. 2. (Color online)

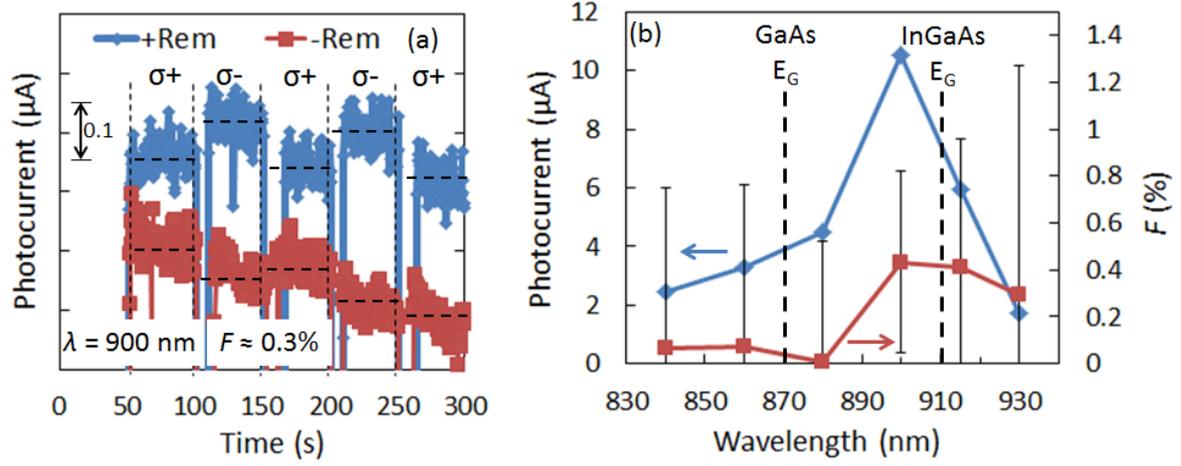

Fig. 3. (Color online)

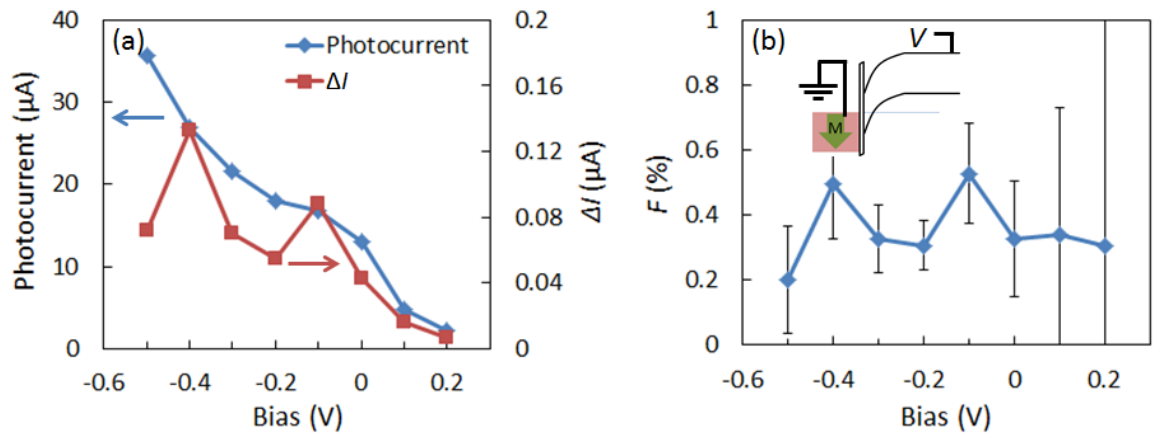

Fig. 4. (Color online)

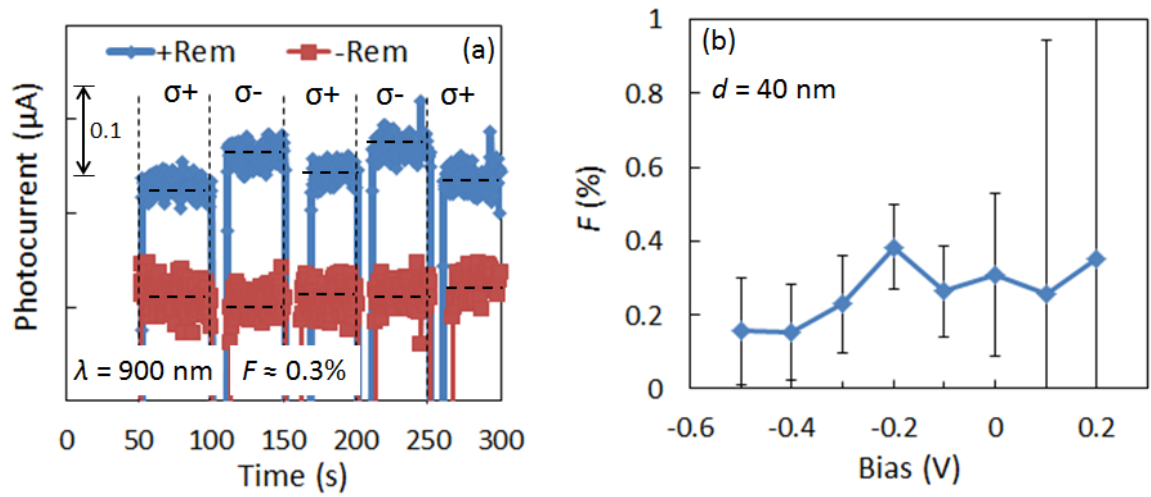

Fig. 5. (Color online)

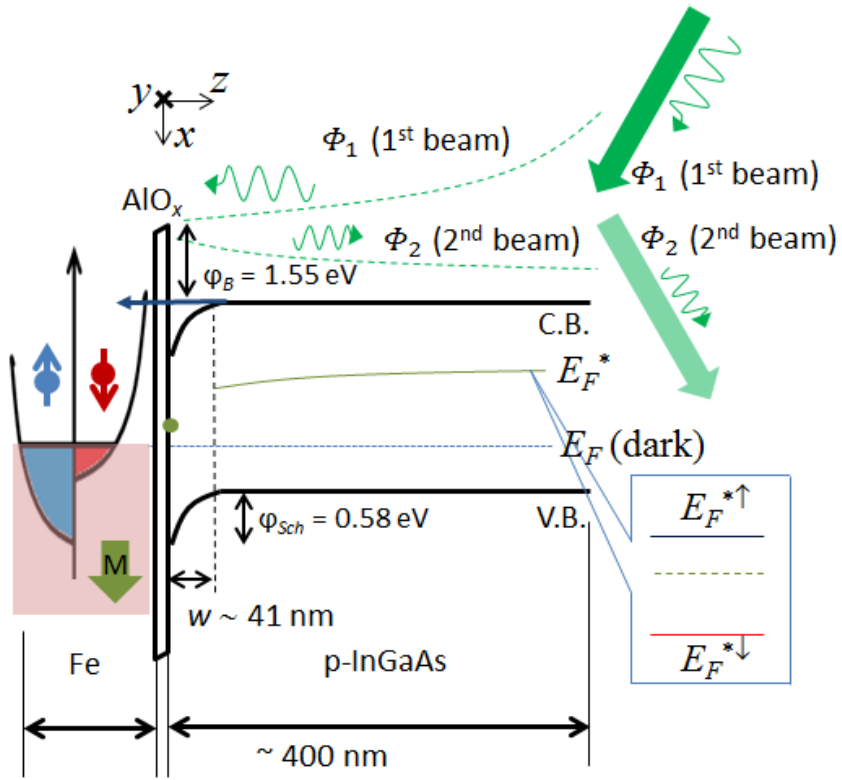

Fig. 6. (Color online)

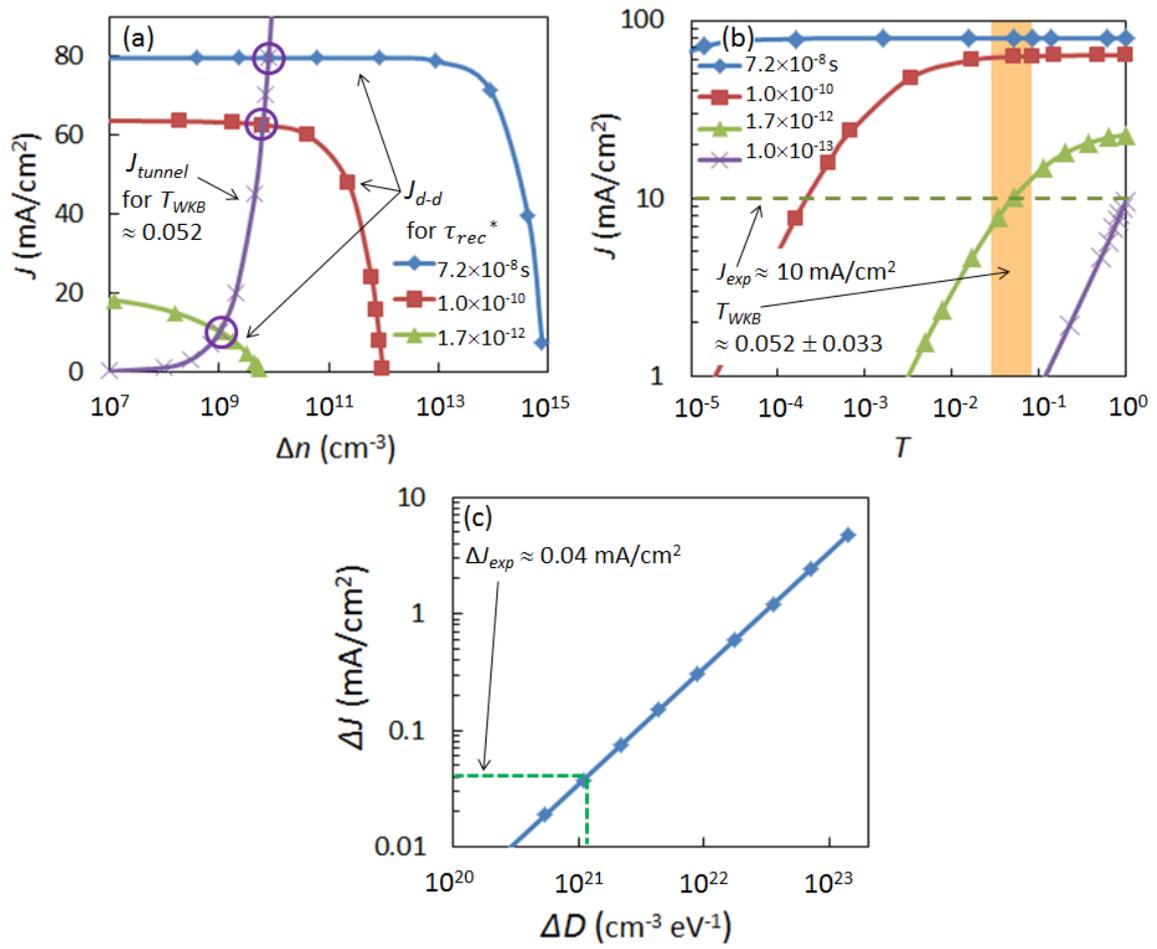

Fig. 7. (Color online)

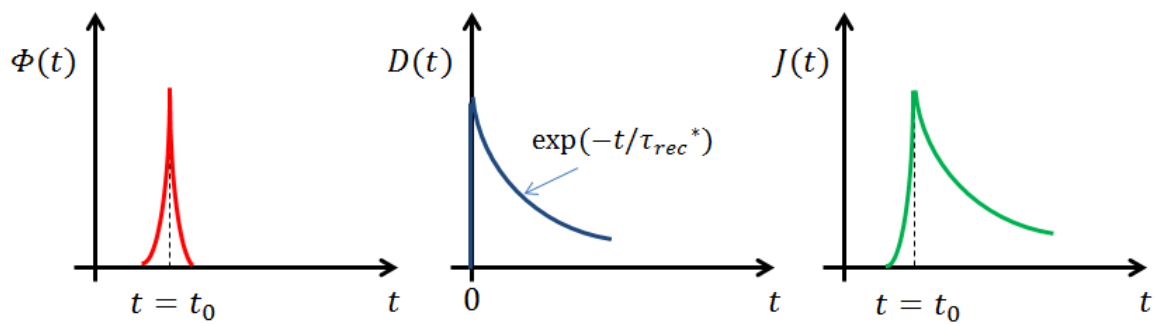

Fig. A.1 (Color online)